# Image denoising through bivariate shrinkage function in framelet domain

**Hamid Reza Shahdoosti[1,*]**

[1] Department of Electrical Engineering, Hamedan University of Technology, Hamedan, Iran
email address: h.doosti@hut.ac.ir

**ABSTRACT**

Denoising of coefficients in a sparse domain (e.g. wavelet) has been researched extensively because of its simplicity and effectiveness. Literature mainly has focused on designing the best global threshold. However, this paper proposes a new denoising method using bivariate shrinkage function in framelet domain. In the proposed method, maximum aposteriori probability is used for estimate of the denoised coefficient and non-Gaussian bivariate function is applied to model the statistics of framelet coefficients. For every framelet coefficient, there is a corresponding threshold depending on the local statistics of framelet coefficients.

Experimental results show that using bivariate shrinkage function in framelet domain yields significantly superior image quality and higher PSNR than some well-known denoising methods.

## 1. INTRODUCTION

Disturbing noise in an imaging system can be either additive or multiplicative. The Gaussian additive white noise (AWGN) has a frequency spectrum which is not only continuous but also uniform over a specified frequency band. This type of noise is spatially uncorrelated. In addition, the noise for each pixel is both independent and identically distributed (iid). One of the main sources of the Gaussian random noise is the operational amplifier and its resistive circuits [1]. It constitutes a serious limitation on the image resolution. Denoising aims at eliminating noise and retaining edges and textures of the images [2].

PDE-based methods can diffuse a degraded image in an anisotropic manner to extract and respect the edge geometry, and allow diffusion along but not across the image edges [3], [4]. Most PDEs make use of the Euler-Lagrange equations corresponding to obtain the denoised image. They assume that the images are piecewise constant to model the natural images [5]. In these methods, the energy function is used with prior terms that penalize the similarity between the noisy and denoised image [6]. A famous family of approaches for image denoising involves the spatially varying convolutions. In these approaches, the noise image is convolved with a spatially-varying, local geometry-driven mask [7]. This idea locally models the noisy image with a low-order (usually less than five) polynomial function coefficients of which are obtained by a weighted least-squares regression, and then the coefficients are used to calculate the value of the denoised image. The weights are selected by means of estimates of the local geometry [8] or the difference in the intensity values between neighboring pixels and the one to be denoised [9]. In this way, noise is removed.

Non-local algorithms [10], [11], are based on the fact that natural images usually have blocks in distant regions that are very similar to each other. These methods obtain the filtered image by minimizing a penalty term on the average weighted distance between the image block and all other blocks of the image, in which the weights are decreasing functions of the squared difference between the values of intensity in the blocks. This results in an update rule which can be used as a spatially varying convolution with non-locally image derived masks. NL-Means can also be uses as a minimizer of the Conditional Entropy (CE) of a central pixel intensity given the intensity values in its neighboring pixels [12], [13].

Transform-domain denoising algorithms typically form another group of powerful image denoising methods. In these algorithms, the image block is mapped into an orthonormal basis, e.g. a wavelet [14], shearlet [15], [16], ripplet [17], [18], and dual-contourlet [19] to give a set of coefficients, which are known to be sparse and uncorrelated for natural images. The coefficients with smaller values often correspond to the higher frequency subbands of the image which are usually dominated by noise. To denoise the image, the coefficients with smaller values are modified (usually, by 'hard thresholding' [20]), and the block is reconstructed by applying the inverse transform. This process is iterated for each block. If the non-overlapping blocks are selected, a large amount of artifacts at the block boundaries and ringing artifacts around edges of the image or salient features is provided, which can be eliminated by performing the cycle spinning technique [21].

In this paper, a new denoising method is proposed based on the bivariate shrinkage function and the framelet transform. Maximum aposteriori probability (MAP) is proposed for estimate of the noise-free coefficient and non-Gaussian bivariate function is employed to model the statistics of neighboring framelet coefficients in different scales and spatial locations. In the proposed method, for each coefficient, there is a corresponding threshold which depends on the local statistics of neighboring coefficients. The proposed method is presented in the next section.

## 2. PROPOSED BIVARIATE SHRINKAGE FUNCTION BASED ALGORITHM

One of the most powerful techniques for image denoising is the bivariate threshold function denoising method which is usually used in wavelet domain. Here, this idea is expended into the framelet domain. In this paper, non-Gaussian bivariate function [22] is applied to model the statistics of framelet coefficients. At first, the relationships between finer scale and coarser scale is defined. Then, the framelet coefficients are modeled. Each coefficient depends on a coefficient at the same spatial location in the immediately coarser scale called CS, and each CS depends on coefficients in the same spatial location in the immediately finer scale called FS. Therefore, each CS in the framelet domain has one CS and each CS has four FS [23]. This model can effectively capture the dependence between a framelet coefficient and its CS.

Generally, for a given original noise-free image, if it is corrupted with an additive white Gaussian noise, the degraded image is as follow:

$$u(x, y) = v(x, y) + n(x, y) \tag{1}$$

where $v(x, y)$, $n(x, y)$ and $u(x, y)$ denote the original noise-free image, noise, and the noisy image, respectively. In the framelet the model is formulated as follows:

$$g = f + e \tag{2}$$

where $g$, $f$ and $e$ denote the observed framelet coefficients, the original noise-free framelet coefficients, and the noisy framelet coefficients, respectively. Denoising aims at obtaining an estimate $\hat{f}$ of the original coefficient such that the estimate value $\hat{f}$ is as close as possible to the original noise-free coefficient $f$.

In the proposed method, Maximum Aposteriori Probability (MAP) is employed for estimate of the denoised coefficient $\hat{f}$. So,

$$\hat{f} = \arg\max_{f} p_{f|g}(f | g) \tag{4}$$

One may use Bayes rule to write:

$$p_{f|g}(f | g) = \frac{p_{g|f}(g | f) p_f(f)}{p_g(g)} \tag{5}$$

So, the following equation can be derived:

$$\hat{f} = \arg\max_{f} \left[ \mathrm{Log}(p_e(g - f)) + \mathrm{Log}(p_f(f)) \right] \tag{6}$$

In order to describe the dependency between the CS and FS in the framelet domain, let $f_1$ represent the CS of $f_2$, (note that $f_1$ is the framelet coefficient at the same spatial position as $f_2$ but at the next finer scale). Then: $f = (f_1, f_2)$, $g = (g_1, g_2)$, and $e = (e_1, e_2)$, $f_1$ and $f_2$ are uncorrelated, but not independent. Thus, $g = f + e$ can be written to

$$\begin{cases} g_1 = f_1 + e_1 \\ g_2 = f_2 + e_2 \end{cases} \tag{7}$$

Assuming that the mean and variance of the added noise are 0 and $s_e^2$, respectively, the noise probability density function denoted by $p_e(e)$ is as follows:

$$p_e(e) = \frac{1}{2ps_e^2} \exp\left(-\frac{e_1^2 + e_2^2}{2s_e^2}\right) \tag{8}$$

Based on the distribution of framelet coefficients, one can fit the model by means of the bivariate probability density function $p_f(f)$ according to the symmetric circular probability distribution [24]. Therefore, it can be shown as the following:

$$p_f(f) = \frac{1}{2/3ps^2} \times \exp\left(-\frac{\sqrt{3}}{s}\sqrt{f_1^2 + f_2^2}\right) \tag{9}$$

Substituting Eqs. (8) and (9) into Eq. (6), yields

$$\hat{f} = \arg\max_f \left[ -\frac{(g_1 - f_1)^2 + (g_2 - f_2)^2}{2s_e^2} + \text{Log}(p_f(f)) \right] \tag{10}$$

If $p_f(f)$ is supposed to be convex, the solving process (Eq. (10)) can be converted into solving the following equations:

$$\begin{cases} \dfrac{g_1 - \hat{f}_1}{s_e^2} + \left(-\dfrac{\sqrt{3}\hat{f}_1}{s\sqrt{\hat{f}_1^2 + \hat{f}_2^2}}\right) = 0 \\ \dfrac{g_2 - \hat{f}_2}{s_e^2} \left(-\dfrac{\sqrt{3}\hat{f}_2}{s\sqrt{\hat{f}_1^2 + \hat{f}_2^2}}\right) = 0 \end{cases} \tag{11}$$

where

$$\sqrt{\hat{f}_1^2 + \hat{f}_2^2} = g_1^2 + g_2^2 - \frac{\sqrt{3}s_e^2}{s} \tag{12}$$

With substitution of Eq. (12) into Eq. (11), the MAP estimator of $f_1$ is the following bivariate shrinkage function:

$$\hat{f}_1 = \frac{\left(\sqrt{g_1^2 + g_2^2} - \left(\dfrac{\sqrt{3}s_e^2}{s}\right)\right)_+}{\sqrt{g_1^2 + g_2^2}} \times g_1 \tag{13}$$

From this equation, one can see that for every framelet coefficient, there is a corresponding threshold, which depends not only on the CS coefficient but also on the FS coefficient.

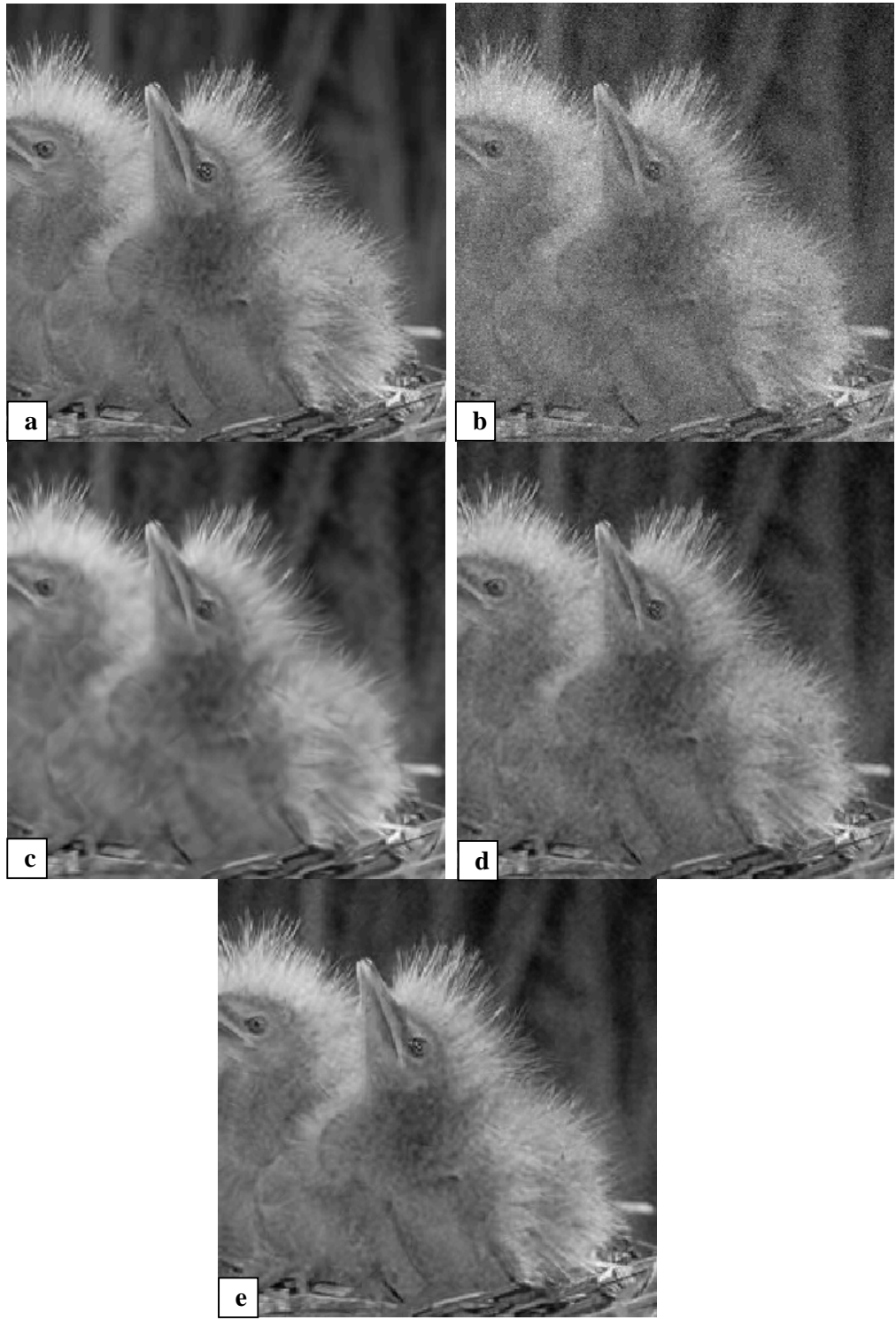

**Fig. 1**. (a) The noise free image. (b) The noisy image ($s=35$). (c) Result of the wavelet-cycle spinning [25]. (d) Result of the curvelet with multi-level threshold [26]. (e) Result of the proposed method.

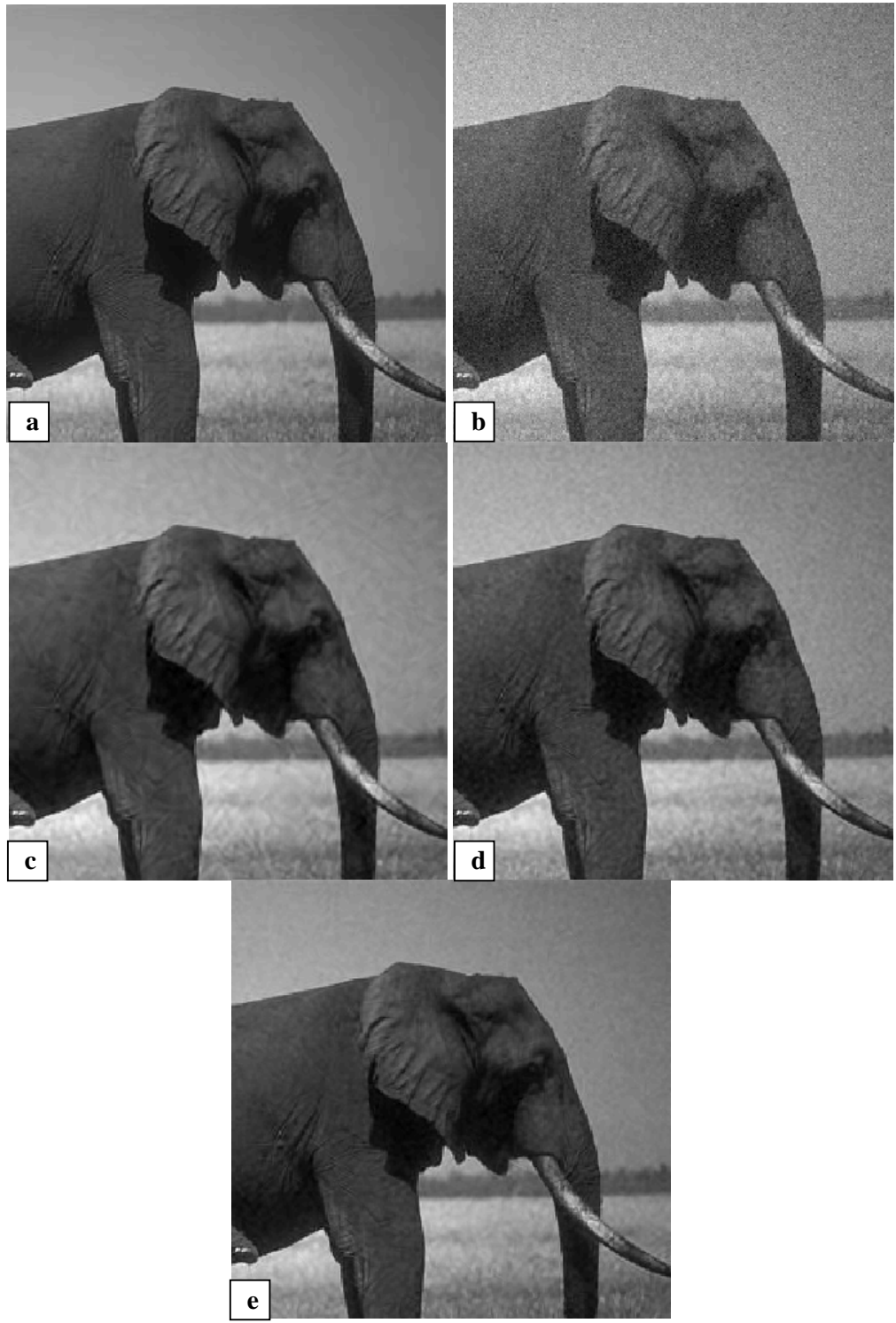

**Fig. 2**. (a) The noise free image. (b) The noisy image ($s = 35$). (c) Result of the wavelet-cycle spinning [25]. (d) Result of the curvelet with multi-level threshold [26]. (e) Result of the proposed method.

3. NUMERICAL EXPERIMENTS

The proposed denoising algorithm using the framelet transform is compared to other effective methods in the literature, namely wavelet-cycle spinning [25], and curvelet with multi-level threshold [26]. Two high resolution of size 256×256 pixels from the Berkeley dataset are chosen for the experiments. These images are shown in Fig. 1 (a) and Fig. 2 (a). The simulated noise images are shown in Fig. 1 (b) and Fig. 2 (b), in which the standard deviation of noise i.e. $s$ is equal to 35. The denoised results obtained by the wavelet-cycle spinning are shown in Fig. 1 (c) and Fig. 2 (c), those obtained by the curvelet with multi-level threshold are shown in Fig. 1 (d) and Fig. 2 (d), and those obtained by the proposed method are shown in Fig. 1 (e) and Fig. 2 (e).

The performance of these denoising methods is tested using the PSNR (Peak Signal-to-Noise Ratio) measure [27-29]. The PSNR values of these methods are tabulated in Table 1. In addition, the results of denoising methods is obtained using the SSIM (Structural Similarity) measure. The SSIM values of these methods are tabulated in Table 2.

**Table 1**. PSNR values for the first noisy image.

| $s$ | Proposed method | Wavelet-cycle spinning | Curvelet with multi-level threshold |
|---|---|---|---|
| 10 | **32.59** | 30.24 | 29.54 |
| 30 | **27.73** | 25.13 | 24.38 |
| 50 | **24.02** | 21.81 | 21.22 |
| 70 | **22.60** | 20.35 | 19.87 |
| 100 | **21.06** | 18.63 | 17.90 |

**Table 2**. SSIM values for the second noisy image.

| $s$ | Proposed method | Wavelet-cycle spinning | Curvelet with multi-level threshold |
|---|---|---|---|
| 10 | **0.96** | 0.85 | 0.86 |
| 30 | **0.89** | 0.78 | 0.79 |
| 50 | **0.81** | 0.69 | 0.71 |
| 70 | **0.74** | 0.60 | 0.65 |
| 100 | **0.69** | 0.55 | 0.61 |

From Table 1, one can find that the proposed method has outstanding results in comparison to the other denoising methods, after considering the relation between the FS and its CS in the framelet domain. One can observe that the proposed method shows higher PSNR gains over other methods (about 2.5dB on average). From Table 2, one can find that the proposed method has outstanding results in comparison to the other implemented methods, after taking into account the relation between the finer scale and coarser scale in the framelet domain. Again, one can observe that the proposed method shows higher SSIM gains over other implemented methods (about 0.1 on average).

Based on the results shown in Figs. 1 (c)-(e) and 2 (c)-(e) and those reported in tables 1 and 2, the proposed method show substantial improvement over other competitors both in visual quality and objective quality using PSNR and SSIM.

4. CONCLUSION

A simple and effective adaptive method was proposed for image denoising through bivariate shrinkage function in framelet domain. The adaptivity is based on the fact that for every framelet coefficient, there is a corresponding threshold, which depends not only on the CS coefficient but also on the FS coefficient. The proposed method made use of the maximum aposteriori probability to estimate the denoised coefficient and non-Gaussian bivariate function to model the statistics of framelet coefficients. As it was shown in this paper, adapting the threshold values to the statistics of framelet coefficients enables us to maintain much of the texture and edge details, while suppressing most of the noise in non-textured regions. This result may be hard to achieve by using a general threshold. The results of experiments show substantial improvement over the other competing algorithms both in visual quality and objective quality e.g. PSNR and SSIM.